\documentclass[twocolumn, prl,aps]{revtex4-1}
\usepackage{graphicx,bm,times}
\usepackage{tabularx} 
\usepackage{lipsum}
\usepackage{amsmath}
\usepackage{amssymb}
\usepackage{gensymb}
\usepackage{bbold}
\usepackage{color,soul, mathrsfs}
 \usepackage{relsize}

\usepackage[]{natbib}

\begin{document}

\title{Beating the Shot-Noise Limit with Sources of Partially-Distinguishable Photons}
\author{Patrick M. Birchall}
\email[]{P.Birchall@Bristol.ac.uk}
\author{Javier Sabines-Chesterking}
\author{Jeremy L. O'Brien}
\author{Hugo Cable}
\author{Jonathan C. F. Matthews}
\email[]{Jonathan.Matthews@Bristol.ac.uk}

\affiliation{Quantum Engineering Technology Labs, H. H. Wills Physics Laboratory and Department of Electrical \& Electronic Engineering, University of Bristol, BS8 1FD, United Kingdom.}

\begin{abstract}
Quantum metrology promises high-precision measurements beyond the capability of any classical techniques, and has the potential to be integral to investigative techniques. However, all sensors must tolerate imperfections if they are to be practical.  Here we show that photons with perfectly overlapped modes, which are therefore fully indistinguishable, are not required for quantum-enhanced measurement, and that partially-distinguishable photons do not have to be engineered to mitigate the adverse effects of distinguishability. We quantify the effect of distinguishability on quantum metrology experiments, and report results of an experiment to verify that two- and four-photon states containing partially-distinguishable photons can achieve quantum-enhanced sensitivity with low-visibility quantum interference. This demonstrates that sources producing photons with mixed spectral states can be readily utilized for quantum metrology.

\end{abstract}

\date{December 30, 2016}
\maketitle
In an ideal scenario, the use of non-classical states of $N$ photons to measure an optical phase $\theta$ will enable the scaling of precision to be increased beyond the shot-noise limit (SNL)---$\delta\theta \propto 1/\sqrt{N}$---to the fundamental Heisenberg limit---$\delta\theta\propto1/N$ \cite{giovannetti2004}. However, a real sensor will operate in non-ideal conditions with non-ideal parameters. Effects such as photon loss and phase diffusion have been shown to remove much of the advantage offered by non-classical techniques, so that the SNL can only be beaten by a constant factor \cite{knysh2011, demkowicz2012}, and this motives further study of imperfections in quantum metrology \cite{giovannetti2011}. Proposals for photonic quantum metrology typically exploit quantum interference of photons which have exactly the same parameters and are therefore perfectly indistinguishable. Achieving this indistinguishability is a major technical challenge in practice, in particular for immature and developing photon source technology. Here we study the effect on precision measurements using non-classical probe states that contain partially-distinguishable photons, which leads to degraded quantum interference. We find that despite high levels of distinguishability, it  is still possible to achieve a quantum advantage in interferometry---if this is the only imperfection and provided there is non-zero indistinguishability, Heisenberg scaling is still achieved. This classifies the effect of distinguishability as separate to those of optical loss and phase diffusion. We have performed a proof-of-principle experiment to observe the quantum advantage that can be achieved with 2- and 4-photon probe states with varied distinguishability. 

Quantum metrology promises an advantage where high precision is needed whilst minimising probe intensity to avoid damaging the system under investigation \cite{taylor2013, wolfgramm2013}. To become useful outside of specialized settings, developments in quantum metrology must address the optical needs of samples being investigated. Creating states which possess desired optical properties with highly-indistinguishable photons is a technological challenge and has been achieved only at a handful of wavelengths \cite{aboussouan2010, harada2011, fulconis2007, harder2013}. However, systems which could use only partially-distinguishable photons would have a greatly-reduced technological challenge associated with creating the photon source. Such systems could dramatically increase the early application of photon sources in spectral regions where highly-indistinguishable photons have not yet been demonstrated \cite{eisaman2011}. Furthermore, schemes based on low-visibility quantum interference could potentially benefit from reduced loss and increased brightness by removing spectral filters.

Interference between two indistinguishable photons, known as Hong-Ou-Mandel (HOM) interference, is now the standard experiment used to quantify the indistinguishability of photons. This is important since HOM interference is the origin of supra-classical performance central to many proposed quantum-optical technologies and is integral to quantum metrology schemes, linear-optical quantum computers, quantum communication and the first loophole-free Bell-inequality violation \cite{holland1993, knill2001, gisin2007, sangouard2011, hensen2015}. The importance of HOM interference has recently inspired many works on photon distinguishability, present in an input state, and its effect on non-universal quantum computers \cite{rohde2015, shchesnovich2015, tichy2014, spagnolo2014, carolan2014, tillmann2014}. In a metrological setting, studies have focused on the effect of photons becoming distinguishable within an interferometer \cite{demkowicz2012} which removes Heisenberg limited scaling. Photonic devices have been reported with fidelities of $\simeq\!100\%$ which indicates that distinguishability already present before the state enters the interferometer can be the dominant effect \cite{laing2010, carolan2015}. 

In Ref.~\cite{jachura2015} a novel approach was presented which utilised two photons with a carefully-engineered additional degree of freedom, and measurement of this additional degree of freedom mitigated the effects of distinguishability on metrological schemes. In this letter we investigate the effect of distinguishability in a situation wherein photons are measured with detectors that resolve only the path the photons are in, and not information about additional degrees of freedom which the photons have. Positive results which apply to this scenario would show that interferometers and detectors do not need to be modified in order to operate using partially-distinguishabile photons. Indeed, the phase sensitivity will be shown here to degrade gently with increasing distinguishability, and precision scaling with increasing photon number remains proportional to the Heisenberg limit. We experimentally observed this degradation by controlling distinguishability between pairs of photons and pairs of bi-photons via a temporal delay. These photons were found to exhibit supra-classical phase sensitivity despite their distinguishability. 

We use the Fisher information to quantify the sensitivity of interference fringes to small changes of the unknown phase, $\theta$, due to its relation with achievable precision in the seminal Cram\'{e}r-Rao bound: $1/\delta^2\theta \leq F$, where $\delta^2\theta$ is the variance of an unbiased estimator of $\theta$ and $F$ is the Fisher information, a function of the probabilities associated with different measurement outcomes \cite{van2000}. This bound can be saturated asymptotically by a large number of measurements, and therefore $F$ well characterises the achievable precision of a scheme. We consider photon-number-counting measurements which are described by the set of projectors $\{\mathbf{E}_{n_1n_2}\}$ such that $\mathbf{E}_{n_1n_2}$ projects onto the subspace containing all states with $n_1$ photons in one path and $n_2$ in the other.

\begin{figure}
\includegraphics[width=1\columnwidth]{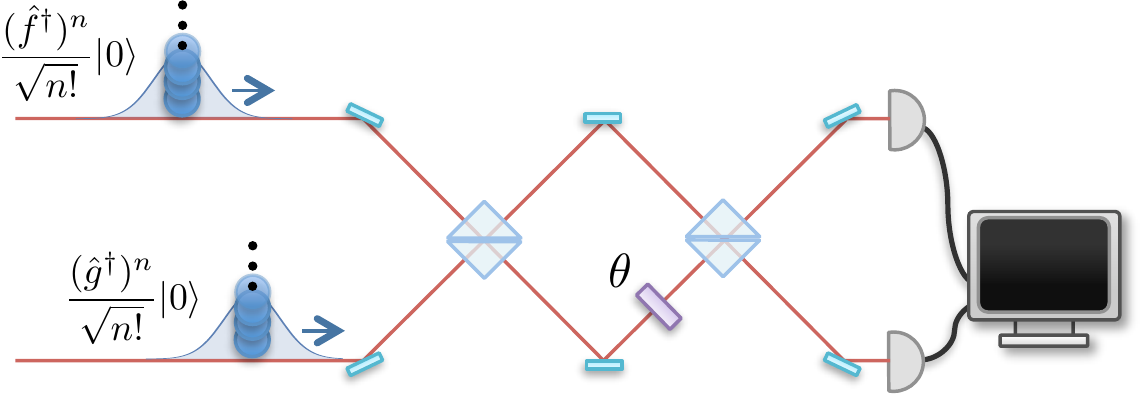}
\caption{\textbf{Mode-mismatch introducing photon distinguishability:} The probe state shown, $|n\rangle_{f,1}|n\rangle_{g,2}$, enters an Mach-Zehnder interferometer before being measured with photon number-counting measurements. The overlap of the temporal modes governs how much quantum interference occurs at the beamsplitters.}
\label{main_scheme}
\end{figure}
To explore the role of photon distinguishability in metrological schemes, we consider distinguishability introduced by mode-mismatch. 
An example of a scheme using a probe state with partially-distinguishable photons is displayed in Fig.~\ref{main_scheme}. Here the probe state is a $2n$-photon dual-Fock state $({\hat{f}}_{1}^{\dagger}{\hat{g}}_{2}^{\dagger})^n/n! |\mathbf{0}\rangle = |n\rangle_{f,1}|n\rangle_{g,2}$ where ${\hat{f}}_{1}^{\dagger}=\int d\omega f(\omega)a_{1}^{\dagger}(\omega)$ and ${\hat{g}}_{2}^{\dagger} =\int d\omega g(\omega)a_{2}^{\dagger}(\omega)$ are boson creation operators for modes $f_1$ and $g_2$. If the temporal modes $f$ and $g$ are mismatched $f\neq g$, then imperfect quantum interference occurs. Output statistics for number-counting measurements can be calculated by changing the basis of the probe state such that it is expressed in terms of modes with well-defined interference relations \cite{rohde2007}. 

Re-expressing the temporal mode $g$ as a linear combination of $f$ and an orthogonal temporal mode, $f_{\bot}\propto g-\langle g,f\rangle f$, the input state becomes:
\begin{equation}
\begin{split}
  |n\rangle_{f,1}|n\rangle_{g,2} =&  \\
    |n\rangle_{f,1}\sum_{k=0}^{n} &\sqrt{\binom{n}{n-k}\mathcal{I}^{n-k}(1-\mathcal{I})^{k}} |n-k\rangle_{f,2}|k\rangle_{f_{\bot},2}
 \end{split}
\end{equation}
where $\mathcal{I}=|\langle f, g\rangle|^2$ is the overlap of the functions $f$ and $g$ and serves as a measure of indistinguishability \cite{ra2013}. For $\mathcal{I}=0$ the photons will be uncorrelated as quantum interference will not occur. Conversely for $\mathcal{I}=1$ the photons will undergo maximal quantum interference. $\mathcal{I}^{\prime}$ is a simple parameterisation of distinguishability which transitions between quantum and classical measurement statistics.
An interferometer acting on both temporal modes performs the operation $e^{i\hat H \theta} =  \mathbb{1} +  i\theta \hat H + \mathcal{O}(\theta^2) $ with $\hat H=i\hat{f}^{\dagger}_1\hat{f}_2 - i\hat{f}_{\bot1}^{\dagger}\hat{f}_{\bot2} + \text{h.c.}$ generating an orthogonal transformation. After transforming the two-photon input state, the probabilities for different detection outcomes are:

\begin{equation}
\begin{split}
p\left(n,n |\theta\right) &=1- \frac{n + \mathcal{I} \,n^2}{2} \theta^2 + \mathcal{O}(\theta^4),\\
p\left(n\pm1,n\mp1|\theta \right) &= \frac{n+ \mathcal{I} \,n^2}{2}\theta^2 + \mathcal{O}(\theta^4).
\end{split}
\end{equation}
This allows us to calculate the phase estimation capabilities of this state by computing the Fisher information:
\begin{equation}
\begin{split}
F &\equiv \sum_{r} \left(\frac{\partial p(r|\theta)}{\partial \theta}\right)^{2}\frac{1}{p(r|\theta)}\\
&= 2(n+\mathcal{I}\,n^2) + \mathcal{O}(\theta^2).
\end{split} \label{fisher}
\end{equation}
For comparison, the SNL for this setup is $F=2n$, therefore if the photons have any indistinguishability ($\mathcal{I} \neq 0$) then the shot-noise limit can be surpassed around $\theta = 0$. Additionally, sensitivity scales quadratically with increasing photon number i.e. proportional to the fundamental Heisenberg limit. If $\theta$ is not near zero then an adaptively controlled phase can be used to counteract $\theta$ producing the sensitivity obtained near $\theta = 0$ \cite{xiang2011}. This point on the interference fringe has been highlighted as a loss-resistant part, retaining the most phase sensitivity in the presence of balanced loss across the arms of the interferometer \cite{datta2011}. In Ref.~\cite{jachura2015}, an alternative approach has been presented to mitigate the effects of distinguishability by using spatially-engineered photons and detectors which resolve the positions of each photon. This approach requires increasing distinguishability in order to mitigate its adverse effects.

 \begin{figure*}
  \includegraphics[width=0.95\textwidth]{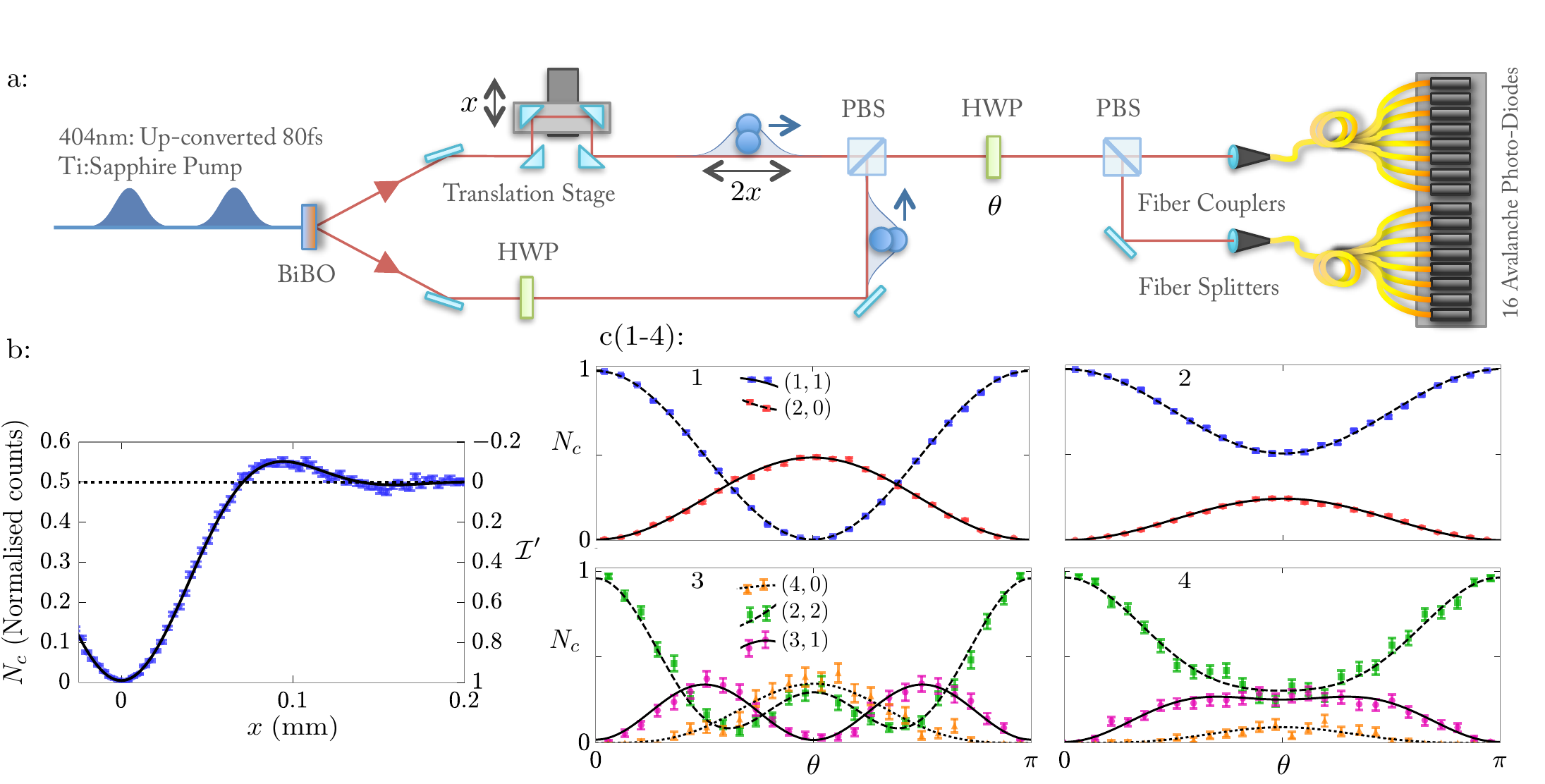}
  \caption{\textbf{Experimental setup and example two- and four-photon interference fringes.} \textbf{a:} Type-I SPDC produces degenerate horizontally-polarised  photon pairs and bi-photon pairs in two spatial modes. One spatial mode is passed through a translation stage to control the mismatch between interfering modes. The other spatial mode passes through a half-wave plate at $45\degree$ rotating horizontally-polarised photons into vertically polarised ones such that the two spatial modes can be combined on a Polarising Beam-Splitter (PBS) and interfered at a HWP. The spatially-multiplexed detection system described in Ref.~\cite{matthews2013}, detects different states with different efficiencies, because of this the raw counts were adjusted by a multiplicative factor to estimate the statistics of a number counting system with uniform detection efficiencies \cite{sperling2012true, achilles2003}. \textbf{b:} Observed photon statistics used to measure indistinguishability as a function of delay $\mathcal{I}^{\prime}(x)$. \textbf{c:} Two- c(1,2) and four-photon c(3,4) interference fringes obtained with high- c(1,3) and low- c(2,4) visibility quantum interference. Error bars are calculated assuming the count rates are given by Poisson statistics. For clarity, only one from each pair of equally probable detection patterns, $\{(n_a,n_b),(n_b,n_a)\}$, are shown.}
  \label{setup}
\end{figure*}

To verify the usefulness of states with partially-distinguishable photons, we performed an experiment using two- and four-photon states post-selected from a Type-I spontaneous parametric down-conversion (SPDC) process. States were post-selected by recording only detection events with the desired number of photons. This allowed us to isolate the effects of distinguishability since the effect of optical loss is also nullified by this procedure. The experimental setup is shown in Fig.~\ref{setup}. Differently from the example above, the two-photon state that Type-I SPDC produces is spectrally entangled $|\psi_{2}\rangle= \iint d\omega_{1}d\omega_{2} \Phi(\omega_{1},\omega_{2})a_{1}^{\dagger}(\omega_1)a_{2}^{\dagger}(\omega_2)|\mathbf{0}\rangle$. Therefore we use a generalised measure of indistinguishability, or exchange symmetry, $\mathcal{I}^{\prime} = \iint d\omega_{1}d\omega_{2}\Phi(\omega_1,\omega_2)\overline{\Phi(\omega_2,\omega_1)}$ which reduces to $|\langle f,g\rangle|^{2}$ for unentangled photons i.e. $\Phi(\omega_1,\omega_2) = f(\omega_1)g(\omega_2)$ \cite{ou2007}. $\mathcal{I}^{\prime}$ can be measured experimentally by observing the coincidence probability after a 50:50 beamsplitter, $p_{\text{H}}$, as in the Hong-Ou-Mandel experiment  \cite{hong1987, ou2007}: $\mathcal{I}^{\prime} = 1-2p_{\text{H}}$ and takes the place of $\mathcal{I}$ in Eq.~(\ref{fisher}) when $n=1$. For a two-photon state created by two single-photon sources separability is guaranteed ensuring $\mathcal{I}^{\prime}$ is positive and supra-classical precision will be achieved. We note that regardless of the spectral structure of such a two-photon state, be it pure or mixed, a single exchange-symmetry parameter, as given by $p_\text{H}$, will completely determine the output statistics after a linear network \cite{adamson2007}. Therefore the two-photon state from Type-I SPDC will give results indicative of all other photon sources with the same $p_\text{H}$. 

The Fisher information of the four-photon data is not determined by $\mathcal{I}^\prime$; however, for some values of $x$, it is determined by the four-photon equivalent of a HOM dip. The post-selected four-photon state entering an interferometer may be expressed as $(2+2\Lambda_4)^{-1/2}\left(\sum_{i} \lambda_i \hat{a}_{1}^\dagger[f_i]\,\hat{a}_{2}^\dagger[g_i]\right)^2|\mathbf{0}\rangle$ with $\Lambda_4 = \sum_i \lambda_i^4$ and $ \sum_i \lambda_i^2=1$, and with two sets of orthonormal spectral functions $\{f_i\}$ and $\{g_i\}$  \cite{law2000}. When $x=0$, due to the symmetry of Type-I SPDC under exchange of paths, the two sets of functions are the same, with $\langle f_i | g_j \rangle = \delta_{i,j}$ \cite{mauerer2009}. Whereas, when $x$ is large $\langle f_i | g_j \rangle = 0$. When $x=0$, the value of $\Lambda_4\in [0,1]$ dictates how often there is genuine four-photon interference and how often there are simply two pairs of mutually indistinguishable photons. By determining $\Lambda_4$ the Fisher information obtained when $x = 0$ ($\mathcal{I}^\prime = 1$) and when $x$ is large $(\mathcal{I}^\prime=0)$ can be predicted as described in appendix C.

We use a half-wave plate (HWP) as a polarisation interferometer as displayed in Fig.~\ref{setup}a. The indistinguishability parameter $\mathcal{I}^{\prime}(x)$ is controlled by delaying one polarisation mode by a distance $x$ with a translation stage prior to the interferometer. Our experiment uses an 80fs pump pulse, and momentum conservation dictates that the down-conversion process results in a theoretical indistinguishability of \cite{ou2007}: $ q(x) =  2\Gamma^{-1}(1/4)\int dy \exp(-y^4)\exp(-iyx/\sigma)$, where $\Gamma$ is the Gamma function \cite{rainville1960special} and $\sigma$ is a constant dependent on the properties of the crystal and pump laser used for down-conversion. We obtained $\mathcal{I}^{\prime}(x)$ experimentally by fitting $a+b\,q(x)$, to normalised coincidence rates shown in Fig.~\ref{setup}a, by allowing $a,b$ and $\sigma$ to vary. At various different values of $x$ the phase of the interferometer was scanned over the range $[0,2\pi)$ obtaining interference fringes with different levels of quantum interference shown in Fig.~\ref{setup}c(1-4). The Fisher information was obtained from fitted curves of $p(r|\theta)$ to normalised interference fringes over $\theta$. See appendix A for the fitting procedures. 

 \begin{figure}
  \includegraphics[width=0.9\columnwidth]{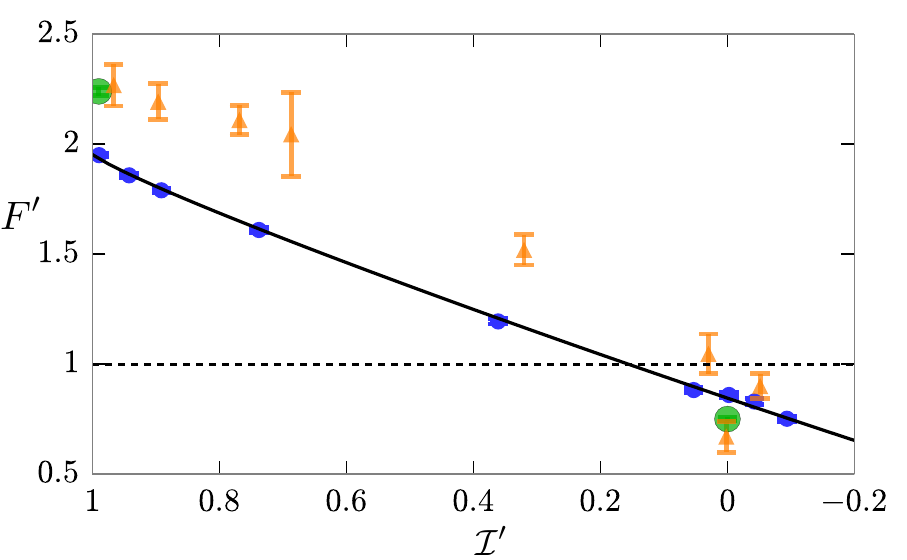}
  \caption{\textbf{Estimates of the Fisher information per photon:} $F^{\prime}\equiv F/(2n)$. Values achieved with two (four) photons are shown in blue (orange). Non-linear relation between $F^{\prime}$ and $I^\prime$ is due to background noise and is modeled in appendix B. Green points are estimates of $F^\prime$ we expect for the four-photon state when $\mathcal{I}^\prime=1$ and when $\mathcal{I}^\prime=0$ as described in appendix C. Error bars arise from the statistical uncertainty in the counts and are found by Monte-Carlo simulation of experimental data followed by the same analysis as the real data which provides a spread of estimates of the Fisher information for each fringe.}
  \label{final_plot}
\end{figure}

The results for Fisher information are shown in Fig.~\ref{final_plot} for both the two- and four-photon input states. The two-photon data demonstrates an approximately-linear degradation of the quantum-enhanced sensitivity in line with the theoretical prediction above. The slight deviation from the linear degradation is due to background noise, which we do not correct for, as modeled in appendix B. The effect of increased background noise is to shift the point of highest sensitivity further away from $\theta = 0$ in addition to lowering the sensitivity of the fringes. For the four-photon input state, since $\mathcal{I}^{\prime}$ by itself does not infer a value for the Fisher information, we use an experimentally-determined estimate of $\Lambda_4 = 0.480\pm0.005$ to estimate the Fisher information we expect around $x = 0$, and for a large $x$, as is shown by green points in Fig.~\ref{final_plot} (appendix C describes how we estimate $\Lambda_4$ which is similar to the method of Ref.~\cite{nagata2007}).  Fig.~\ref{final_plot} shows the sensitivity of the four-photon fringes decays from the predicted value when $\mathcal{I}^\prime=1$ approximately down to the predicted value for $\mathcal{I}^\prime=0$ as expected. As we cannot measure the function $\Phi(\omega_1,\omega_2)$ we can only predict the value of $F$ in the extremal cases of small and large $x$. We note that our four-photon input states give a higher precision that previously reported experiments using similar states \cite{xiang2011}. In order for supra-classical values of Fisher information to be extracted contrast must remain high around $\theta=0$. This is apparent in the fringes we obtained experimentally as shown in Fig.~\ref{setup}c.

We have shown that the effects of photon distinguishabiltiy have a distinct nature from errors considered previously, as Heisenberg-limited scaling remains with partially-distinguishable photons. Surprisingly, any amount of quantum interference can grant a quantum advantage in the absence of any other imperfections. The experiment we performed allowed us to see the gentle degradation of precision in agreement with our theoretical prediction. We conclude that sources of highly-indistinguishable photons are not necessary to gain a quantum advantage in metrology. Our results reduce technical difficulties associated with making photon sources useful for metrology. Using novel photon sources such as integrated sources \cite{harada2011}, lattice defect sources \cite{castelletto2014}, atomic sources \cite{nisbet2011}, heralded sources \cite{ma2011experimental,thomas2011} using fast switching, or quantum memories \cite{collins2013, simon2010} and quantum dots \cite{gazzano2013} may enable quantum enhanced measurements to be performed in new spectral regions. 

\textbf{Acknowledgments:} The authors would like to thank J. D. A. Meinecke, T. Ono and D. Bonneau for helpful discussions. This work was supported by EPSRC, ERC, PICQUE, BBOI, US Army Research Office (ARO) Grant No. W911NF-14-1-0133, U.S. Air Force Office of Scientific Research (AFOSR) and the Centre for Nanoscience and Quantum Information (NSQI). J.L.O’B. acknowledges a Royal Society Wolfson Merit Award and a Royal Academy of Engineering Chair in Emerging Technologies. J.C.F.M. and J.L.O'B acknowledge fellowship support from the Engineering and Physical Sciences Research Council (EPSRC, UK). J.S.C. acknowledges support from the University of Bristol.


%

\section{Appendix}
\subsection{A:  Function fitting}

To obtain the estimates of the Fisher information, for the post-selected two-photon input state as plotted in Fig.~3 of the main text, the probability functions for each of the outcomes were first estimated. Theoretically, the probability functions are:
\begin{equation} \label{two_probs}
\begin{split}
p\left(|\Delta| = 1|\theta \right) &= \frac{1 + \mathcal{I}^\prime}{4}\left[1- \cos(2 \theta)\right], \\
p\left(|\Delta| = 0|\theta \right) &=\frac{1}{4}\left[ 3 - \mathcal{I}^\prime + (1 + \mathcal{I}^\prime) \cos(2 \theta) \right],
\end{split}
\end{equation}
where $\Delta = n_1-n_2$. The detection efficiency of each outcome $\eta_{\Delta}$ will affect the observed statistics so we correct for this in order to estimate the probability functions which would be observed with ideal detection efficiency. 
These functions are fitted to our data by allowing the Fourier coefficients; $\mathbf{c}$ to vary. We fit probability functions with positive and negative $\Delta$ separately. Maximum-likelihood estimation is used to find these coefficients. 

Firstly, estimates of ideal detection rates are obtained by dividing observed detection rates by the intrinsic detection inefficiency given by the combinatorics of a multiplexed detection system \cite{sperling2012true, matthews2013}. To account for the unequal count rates across each interference fringe we fit the functions in Eq.~(\ref{two_probs}) to normalized count rates, that is the counts of a specific outcome divided by the total number of counts at this position. The likelihood function to be maximized is:
\begin{equation}
\begin{split}
\mathcal{L}(\mathbf{c}_{\Delta}|\mathbf{x}_{\Delta})&= \prod_{\theta} p(x_\Delta^\theta|\lambda_{\Delta}^\theta),\\
\lambda_{\Delta}^\theta& = \lambda_t^\theta \times p(\Delta|\theta) \times \eta_\Delta
\end{split}
\end{equation}
where $x_\Delta^\theta$ are the observed number of counts and $\mathbf{x}_{\Delta}$ is a vector of such over $\theta$, $\lambda_{\Delta}^\theta$ is an estimate for the expected number of observed counts of $\Delta$ type given $\mathbf{c}_{\Delta}$, $\lambda^{\theta}_{t} = \sum_{\Delta} x_{\Delta}^{\theta}/\eta_{\Delta}$ is a maximum likelihood estimate for the total number of events an ideal detector would record at this value of $\theta$. The Log likelihood function was maximized numerically. Many random initial guesses of $\mathbf{c}_{\Delta}$ are followed by gradient ascent to try and find the global likelihood maximum.

To find the statistical error associated with our analysis, this curve-fitting procedure was conducted many times with counts which has been simulated base on the probability functions fitted to the real data. The standard error of these repeats was taken to be the statistical error. 

The probability functions for the four-photon interference fringes were granted an additional Fourier term as they should contain oscillations with double the frequency over the two-photon state. These coefficients do not have an expression in terms of $\mathcal{I}^{\prime}$. The same procedure was carried out for the post-selected four-photon input state to perform curve fitting and error analysis. 

\subsection{B: Predicted $F$ vs $\mathcal{I^\prime}$ for two-photon fringes.}
The dominant source of background noise in the two-photon interference fringes we observed is accidental coincidences between detected photons which did not arise from a SPDC event. We measured the accidental coincidences by recording the coincidences after delays had been added to the output signals of our avalanche photodiodes such that any pairs from an SPDC event would no longer be recorded as a coincidence. This leaves only the rate of coincidences caused by uncorrelated photons. We measured this rate to be $15.6\pm0.4$Hz after correcting for non-uniform detection efficiency. The two-photon count rate when the detectors signals were matched was $1315\pm5$Hz which results in a probability of a count being background of $1.19\pm0.03$\%. These rates were found by integrating over two minutes with the phase of the HWP $0^\circ$ and the intensity of the laser the same as it was when the two-photon interference fringes were recored. The errors in average count rates were found assuming Poisson statistics and have been propagated to the error in background probability by standard error propagation techniques. 

Rather than correct for background noise, we include this effect into our analysis such that the ideal phase dependent two-photon probabilities, $p\big(|\Delta|\,\big|\theta \big)$, are modified to:
\begin{equation}
p_{D}\big(|\Delta|\,\big|\theta \big) = p\big(|\Delta|\,\big|\theta \big)(1-\zeta)+\zeta/2.
\end{equation}
Where $\zeta=1.19\pm0.03$\% is the probability of observing a count due to background noise. Using these probabilities we find the largest Fisher information to be at a different point on the interference fringe, for the two-photon fringes the position with highest phase sensitivity is given by:
\begin{equation}
\theta = \arctan \left( \sqrt[\leftroot{-2}\uproot{2}\mathlarger{4}]
{\frac{\zeta^2-2\zeta}{{\mathcal{I}^\prime}^2(\zeta-1)^2-1 }}\, \right).
\end{equation}
The resulting Fisher information can be found by evaluating Eq.~(3), within the main text, at this point. This gives an algebraic expression for optimal Fisher information:
\begin{equation} \label{optimal_fi}
2+ 2 \mathcal{I}^{\prime}(\zeta-1)^2 +2 \sqrt{\zeta(\zeta - 2)[{\mathcal{I}^{\prime}}^2(\zeta-1)^2-1]}.
\end{equation}
Inserting $\zeta=1.19\%$ into Eq.~(\ref{optimal_fi}) gives the theoretical line displayed in Fig.~3 of the main text.

\subsection{C: Predicted $F$ vs $\mathcal{I^\prime}$ for four-photon fringes.}
To predict the Fisher information of the four-photon state we follow a procedure similar to that of Ref.~\cite{nagata2007}. We experimentally determine properties of the four-photon state and then use this, in combination with background noise rates, to predict the Fisher information. As in appendix B we measured the rate of four-fold coincidences, when the coincidence time bins were matched, to be $2.297\pm0.023$ Hz and when the time bins were mismatched to be $0.065\pm0.010$ Hz. The probability of a four-fold coincidence being background noise is therefore $2.82\pm0.0045\%$.

As stated in the main text, the expression for the four-photon state when $x = 0$ is $\frac{1}{\sqrt{2+2\Lambda_4}}\left(\sum_{i} \lambda_i \hat{a}_{1,i}^\dagger\,\hat{a}_{2,i}^\dagger\right)^2|\mathbf{0}\rangle$ where we have used abbreviated notation $\hat{a}_{1,i}^\dagger \equiv \hat{a}_{1}^\dagger[f_i]$ since there is only one set of spectral functions $\{f_i\}$. Passing this state through a HWP at $22.5\degree$ applies a beamsplitter like transformation: $\hat{a}_{1,i}^\dagger \xrightarrow{U(\pi/4)}(\hat{a}_{1,i}+\hat{a}_{2,i})/\sqrt{2}$, $\hat{a}_{2,i}^\dagger \xrightarrow{U(\pi/4)}(-\hat{a}_{1,i}+\hat{a}_{2,i})/\sqrt{2}$ to the state:
\begin{widetext}
\begin{equation}
\begin{split}
& \frac{1}{\sqrt{2+2\Lambda_4}}\left(\sum_{i} \lambda_i \hat{a}_{1,i}^\dagger\,\hat{a}_{2,i}^\dagger\right)^2|\mathbf{0}\rangle \\
= &\,\,\frac{1}{\sqrt{2+2\Lambda_4}}\left(\sum_i \lambda_i^2\, {{}{\hat{a}_{1,i}^{\dagger}}}^2 {{}\hat{a}_{2,i}^\dagger}^2 +2 \sum_{i>j}   \lambda_i\lambda_j {{}{\hat{a}_{1,i}^{\dagger}}}{{}{\hat{a}_{2,i}^{\dagger}}}{{}{\hat{a}_{1,j}^{\dagger}}}{{}{\hat{a}_{2,j}^{\dagger}}} \right)|\mathbf{0}\rangle \\
&\!\!\xrightarrow{U(\pi/4)}  \,\,  \frac{1}{\sqrt{2+2\Lambda_4}} \left[ \sum_i \lambda_i^2 \left({{}{\hat{a}_{1,i}^{\dagger}}}^4 - 2 {{}{\hat{a}_{1,i}^{\dagger}}}^2 {{}{\hat{a}_{2,i}^{\dagger}}}^2 + {{}{\hat{a}_{2,i}^{\dagger}}}^4 \right) \big/ 4 
+(1/2)  \sum_{i>j} \lambda_i\lambda_j \left( -{{}{\hat{a}_{1,i}^{\dagger}}}^2 + {{}{\hat{a}_{2,i}^{\dagger}}}^2 \right)\left( -{{}{\hat{a}_{1,j}^{\dagger}}}^2 + {{}{\hat{a}_{2,j}^{\dagger}}}^2 \right)  \right]  |\mathbf{0}\rangle \\
= &\, \frac{1}{\sqrt{2+2\Lambda_4}}\! \left[\! \sum_i \lambda_i^2 \!\left(\frac{\sqrt{3}}{\sqrt{2}} |4\rangle_{1,i} \!-\! |2\rangle_{1,i}|2\rangle_{2,i}  \!+\! \frac{\sqrt{3}}{\sqrt{2}} |4\rangle_{2,i}  \right)\!+  \sum_{i>j} \lambda_i \lambda_j  \Big( |2\rangle_{1,i}|2\rangle_{1,j} \!+\! |2\rangle_{1,i}|2\rangle_{1,j} \!-\! |2\rangle_{1,i}|2\rangle_{1,j}\!-\! |2\rangle_{1,i}|2\rangle_{1,j} \Big)\!\right] \\
\end{split}
\end{equation}
\end{widetext}
yielding $p(|\Delta| = 4) = (2\Lambda_4 +1)(2\Lambda_4+2)$ and hence $\Lambda_4 = [2p(|\Delta| = 4)+1]/[2-2p(|\Delta| = 4)]$. Using the four-photon interference fringe when $x=0$, from the fitted functions, we find that $p(|\Delta|=4)=0.6619\pm0.0012$ once the background noise has been accounted for. Therefore $\Lambda_4 = 0.4790\pm0.0053$. Then using this value for $\Lambda_4$, and the level of background noise, we could evaluate an expected Fisher information across $\theta \in [0,\pi)$ when $x=0$ ($\mathcal{I}^\prime=0$) and for large $x$ ($\mathcal{I}^\prime = 1$) to have maximums of $2.246\pm0.039$ and $0.7547\pm0.017$. Errors on these values have been calculated by finding the expected value for Fisher information using the extremal errors for the background noise. The points are plotted as green circles in Fig.~3 of the main text and the error bars are smaller than the points. As can be observed in Fig.~3 these values suitably agree with the experimentally obtained values for Fisher information when $\mathcal{I}^\prime=0$ and when $\mathcal{I}^\prime = 1$.

\end{document}